\documentclass[pss]{wiley2sp} 
\usepackage{amsmath}
\usepackage[squaren]{SIunits}

\begin{document}

\newcommand{\comment}[1]{!!! \textbf{#1} !!!}

\title{Phase separation of multicomponent excitonic Bose-Einstein condensates}

\titlerunning{Bose--Einstein condensation of excitons}

\author{%
  S. Sobkowiak\textsuperscript{\Ast,\textsf{\bfseries 1}},
  D. Semkat\textsuperscript{\textsf{\bfseries 1}},
  H. Stolz\textsuperscript{\textsf{\bfseries 1}},
  Th. Koch\textsuperscript{\textsf{\bfseries 2}},
  H. Fehske\textsuperscript{\textsf{\bfseries 2}}}

\authorrunning{S. Sobkowiak et al.}

\mail{e-mail
  \textsf{siegfried.sobkowiak@uni-rostock.de}, Phone:
  +49-381-4986927, Fax: +49-381-4986922}

\institute{%
  \textsuperscript{1}\,Institut f\"ur Physik, Universit\"at Rostock, 18051 Rostock, Germany\\
  \textsuperscript{2}\,Institut f\"ur Physik, Ernst--Moritz--Arndt--Universit\"at Greifswald, 17489 Greifswald, Germany}

\received{XXXX, revised XXXX, accepted XXXX} 
\published{XXXX} 


\abstract{%
%
%
%
\abstcol{%
For the observation of Bose--Einstein condensation, excitons in cuprous oxide are regarded as promising candidates due to their large binding energy and long lifetime. High particle densities may be achieved by entrapment in a stress induced potential. We consider a multi-component gas of interacting para- and orthoexcitons in}{ cuprous oxide confined in a three-dimensional potential trap. Based on the Hartree--Fock--Bogoliubov theory, we calculate density profiles as well as decay luminescence spectra which exhibit signatures of the separation of the Bose-condensed phases.
  }
}

\maketitle   

\section{Introduction}
The theoretical framework for trapped dilute interacting bosonic gases is well known from the theory of atomic condensates \cite{griffin1996,DGP99,BFB00,PJ08}. First applications to excitonic systems exist, as well \cite{BHB04}. Recent investigations in the framework of a mean-field formalism in local density approximation suggest distinct signatures of a condensate in the decay luminescence spectrum of the thermal excitons \cite{SS10,SSSKF10,siegfrieddiplom}. Works on two-component systems \cite{HS96,BV97,SZC00,VS} have shown that the occurrence of phase separation is closely tied to the proportions of inter- and intra-species interaction strengths. Yet, the description of exciton--exciton interaction is a long-standing problem. Experimental results and theoretical predictions for the interaction strengths vary within an order of magnitude \cite{SC01,BHB04,Brandt07}.
As an example, we present numerical results for the densities and the spatially resolved luminescence spectra of the three component system of excitons in cuprous oxide (Cu$_2$O), i.e., para-, ortho(+)-, and ortho($-$)excitons, trapped in a strain induced potential \cite{SN00}. We show how spectral features may reveal phase separation, thereby yielding a minimum estimate of the relative strength of the mutual interactions.

\section{Multicomponent exciton systems}

\subsection{Thermodynamics}
We consider a $K$-component exciton gas in second quantization, starting from the Hamiltonian in the grand canonical ensemble:
\begin{eqnarray}
\label{eq:Hamiltonallg}
\mathcal{H}&=& \sum_{i=1}^K \int \mathrm{d}^3\mathbf{r} \ \psi^\dagger_i(\mathbf{r},t)\left( -\frac{\hbar^2 \nabla^2}{2 M_i} + V_i(\mathbf{r})-\mu_i \right)  \psi_i(\mathbf{r},t) \nonumber \\
&+&\frac{1}{2}\sum_{i,j=1}^K \int \mathrm{d}^3\mathbf{r} \  h_{ij} \psi^\dagger_i(\mathbf{r},t) \psi^\dagger_j(\mathbf{r},t) \psi_j(\mathbf{r},t) \psi_i(\mathbf{r},t)\;,
\end{eqnarray}
where $V_i$ represents the external (trap) potentials and $\mu_i$ the chemical potentials.
We assume a contact potential for the exciton--exciton interaction, with the intra- and inter-species interaction strengths $h_{ij}=2\pi\hbar^2(M_i^{-1}+M_j^{-1})a^s_{ij}$ given by the respective s-wave scattering lengths $a^s_{ij}$. 

The Bose field operators $\psi_i(\mathbf{r},t)$ are decomposed in the usual fashion, $\psi_i(\mathbf{r},t)=\varPhi_i(\mathbf{r})+\widetilde{\psi}_i(\mathbf{r},t)$, with the condensate wave functions $\varPhi_i(\mathbf{r})=\langle \psi_i(\mathbf{r},t) \rangle=\langle \psi_i(\mathbf{r}) \rangle$ and the operators of the thermal excitons $\widetilde{\psi}_i(\mathbf{r},t)$. The Heisenberg equations of motion $\mathrm{i}\hbar\partial_t \psi_i=[\psi_i,\mathcal{H}]$ result in $2K$ coupled equations (arguments dropped for brevity): the Gross-Pitaevskii equations (GPE) for the condensates,
\begin{eqnarray}
\label{eq:GPE3Kallgerg}
0\! &=&\! \Bigg(\! -\frac{\hbar^2\nabla^2}{2M_i} +V_i-\mu_i + h_{ii} \left ( n_{ii} + \widetilde n_{ii}\right ) +\!\sum_{j\neq i} h_{ij} n_{jj}  \!\Bigg) \varPhi_i \nonumber \\
&& + \, h_{ii} \widetilde{m}_{ii} \varPhi_i^* +\sum_{j\neq i} h_{ij} \Big(   \widetilde{n}_{ji} \varPhi_j  + \widetilde{m}_{ji}  \varPhi^*_j \Big)\;,
\end{eqnarray}
and the equations of motion for the thermal excitons,
\begin{eqnarray}
\label{eq:Beweg3Kallgerg}
\mathrm{i}\hbar\frac{\partial \widetilde{\psi}_i}{\partial t}\! &=&\! \ \Bigg(\! -\frac{\hbar^2\nabla^2}{2M_i} +V_i-\mu_i + 2 h_{ii}  n_{ii} +\!\sum_{j\neq i} h_{ij} n_{jj} \!\Bigg) \widetilde{\psi}_i\nonumber \\ 
&&+ \, h_{ii}m_{ii} \widetilde{\psi}^\dagger_i 
+ \sum_{j\neq i} h_{ij} \Big ( n_{ij} \widetilde{\psi}_j + m_{ij} \widetilde{\psi}^\dagger_j \Big )\;.
\end{eqnarray}
Here $n_{ij}\equiv \varPhi_j^\ast\varPhi_i + \widetilde n_{ij}$, $m_{ij}\equiv \varPhi_j\varPhi_i + \widetilde m_{ij}$, with the averages 
$\widetilde{n}_{ij}=\langle \widetilde{\psi}^\dagger_i \widetilde{\psi}_j \rangle$ and $\widetilde{m}_{ij}=\langle \widetilde{\psi}_i \widetilde{\psi}_j \rangle$.
For simplicity, we neglect all non-diagonal averages, i.e., the last terms on the r.h.s of (\ref{eq:GPE3Kallgerg}) and (\ref{eq:Beweg3Kallgerg}) and subsequently obtain effective one-component equations with mean field contributions from the respective other components.
Because the extension of the potential trap is large compared to the thermal deBroglie wavelength of the excitons, we apply a local density approximation to (\ref{eq:Beweg3Kallgerg}), setting $\nabla^2\to-|\mathbf{k}|^2$ with a wave vector $\mathbf{k}$. For the same reason, we apply the Thomas-Fermi approximation to the GPE, thus neglecting the kinetic energy term in (\ref{eq:GPE3Kallgerg}).
With the above simplifications, Eq.\ (\ref{eq:Beweg3Kallgerg}) is solved by a Bogoliubov transformation 
%
\begin{equation}
\widetilde{\psi}_i=\sum_\mathbf{k}\left[u_i(\mathbf{k}) a_i^{}(\mathbf{k})\mathrm{e}^{-\mathrm{i} E_i(\mathbf{k})t/\hbar}+v_i^*(\mathbf{k}) a_i^{\dag}(\mathbf{k}) \mathrm{e}^{\mathrm{i} E_i(\mathbf{k}) t/\hbar}\right]\,.
\end{equation} 
The densities $n_i^T\equiv\widetilde n_{ii}$ of thermally excited excitons are given by
\begin{eqnarray}
\label{eq:nT3K}
n_{i}^T(\mathbf{r}) &=& \int \frac{\mathrm{d}^3\mathbf{k}}{8\pi^3} \left[ \frac{L_i(\mathbf{k},\mathbf{r})}{E_i(\mathbf{k},\mathbf{r})}\left( n_{B}(E_i(\mathbf{k},\mathbf{r}))+\frac{1}{2} \right)-\frac{1}{2} \right]\nonumber\\ &&\times\Theta\left(E_i(\mathbf{k},\mathbf{r})^2\right) 
\end{eqnarray}
with $n_{B}(E)= \left[\exp(E/k_B T)-1\right]^{-1}$. To guarantee gapless excitation spectra $E_i$, we neglect all anomalous averages $\widetilde m_{ii}$ (Popov approximation) and obtain
\begin{eqnarray}
\label{eq:Energie3k}
E_i(\mathbf{k},\mathbf{r})\! &=&\! \sqrt{L_i(\mathbf{k},\mathbf{r})^2-(h_{ii}n_{i}^c(\mathbf{r}))^2}\,, \\
L_i(\mathbf{k},\mathbf{r})\! &=&\! \frac{\hbar^2k^2}{2M_i} +V_{i}(\mathbf{r})-\mu_i + 2 h_{ii} n_{i}(\mathbf{r}) +\! \sum_{j\neq i} h_{ij} n_{j}(\mathbf{r})\nonumber\,,
\\
\end{eqnarray}
with $n_i^c\equiv|\varPhi_i|^2$ and $n_i\equiv n_{ii}=n_i^T+n_i^c$.

From the simplified GPEs, the condensate densities follow as
\begin{eqnarray}
\label{eq:nc3K}
n_{i}^c(\mathbf{r})=\frac{1}{h_{ii}}\Big(\mu_i-V_{i}(\mathbf{r})
-2h_{ii}n_{i}^T(\mathbf{r})
-\sum\limits_{j\neq i} h_{ij} n_{j}(\mathbf{r})\Big)\,,\;\;\;
\end{eqnarray}
if this expression is non-negative, and $n_{i}^c(\mathbf{r})=0$ otherwise.
Equations (\ref{eq:nT3K}) to (\ref{eq:nc3K}) have to be solved self-consistently. Although they look similar to the one-component case, the coupling between the components appears in $L_i$ and $n_{i}^c$.

\subsection{Luminescence spectrum}
Excitons decay by emitting photons. We apply a local approximation to the emission spectrum, which is determined by the excitonic spectral function $A(\mathbf{k},\omega)$ \cite{shi1994,haug1983}:
\begin{eqnarray}
\label{eqn:spectrum}
I_i(\mathbf{r},\omega) & \propto & 2 \pi |S_i(\mathbf{k}=0)|^2 \delta(\hbar\omega^\prime - \mu_i) n_{i}^c(\mathbf{r}) \\
&+& \sum_{\mathbf{k}\ne 0}|S_i(\mathbf{k})|^2n_{B}(\hbar\omega^\prime - \mu_i)A_i(\mathbf{r},\mathbf{k},\hbar\omega^\prime - \mu_i)\;, \nonumber
\end{eqnarray}
with the exciton-photon coupling $S_i(\mathbf{k})$. The spectral function is given by the Bogoliubov amplitudes $u_i$ and $v_i$, and by the quasiparticle spectrum in (\ref{eq:Energie3k}):
\begin{eqnarray}
A_i(\mathbf{r},\mathbf{k},\omega)&=&2\pi\hbar\Bigg[ u_i^2(\mathbf{k},\mathbf{r}) \delta(\hbar\omega-E_i(\mathbf{k},\mathbf{r})) \\
&&-\,v_i^2(\mathbf{k},\mathbf{r}) \delta(\hbar\omega+E_i(\mathbf{k},\mathbf{r})) \Bigg] \,.\nonumber
\end{eqnarray}
In Cu$_2$O, the decay of orthoexcitons takes place via momentum supplying phonons, such that all exciton states $\mathbf{k}$ participate and $\omega^\prime = \omega - E_{gX}/\hbar -\omega_{\rm phonon}\,$, with $E_{gX}$ being the excitonic band gap. The paraexcitons decay without phonons, i.e., $\omega^\prime = \omega - E_{gX}/\hbar$. Here, energy and momentum conservation only allow for processes, where the wave vectors of excitons and photons are equal, so that $S(\mathbf{k})=S_0\delta(\mathbf{k}-\mathbf{k}_0)$, with $|\mathbf{k}_0|= E_{gX} n/\hbar c$. Therefore, the condensate does not contribute to the zero-phonon process. Despite this, there are indirect signatures of the condensate in the spatially resolved luminescence spectrum of the non-condensed excitons \cite{SS10}.

In a typical experiment, one images a small stripe of width $2\Delta x$ elongated along the $z$-direction onto the 
entrance slit of a spectrograph. Thus, by integrating (\ref{eqn:spectrum}) over the $x$- and $y$-directions perpendicular to $z$, we obtain the spatially resolved spectrum. Moreover, we account for the finite spectral resolution $\Delta$ by convoluting the spectral intensity with a slit function of the shape $\exp\{-(\omega/\Delta)^{2}\}$. For our calculations, we use values of $\Delta=\unit{25}{\micro\electronvolt}$ for the spectral resolution and $\Delta x=\unit{25}{\micro\meter}$ for the entrance slit of the spectrograph being typical for a triple high-resolution spectrograph used in the current experiments which are underway \cite{schwartz2010}.

\section{Results}

We calculated the density distributions and the luminescence spectra of the trapped excitons for three different choices of
the interaction strength between ortho(+)- and ortho($-$)excitons ($h_{+-}$). The interaction strengths are calculated from the scattering lengths given in \cite{SC01} ($h_{++}=h_{--}=0.71 h_{pp}$, $h_{p+}=h_{p-}=0.33 h_{pp}$, and $h_{+-}=1.77 h_{pp}$ with $h_{pp}=\unit{5.4 \times 10^{-4}}{\micro\electronvolt\micro\meter\cubed}$). Taking the value for $h_{+-}$ given by Shumway and Ceperley \cite{SC01} as $h_{SC}$ we chose: (i) $h_{+-}=h_{SC}/5$; (ii) $h_{+-}=h_{SC}/3$; and (iii) $h_{+-}=h_{SC}$ while keeping the others fixed. The used trap potentials $V_i$ are fitted to experimental data taken from \cite{schwartz2010}. The minimum of the paraexciton trap is $\unit{-1981}{\micro\electronvolt}$ and the minimum of the orthoexciton trap is $\unit{-8158}{\micro\electronvolt}$. Note that we neglect the difference in paraexciton and orthoexciton mass due to the $\mathbf{k}$-dependent exchange interaction \cite{Dasbach}. 

\begin{figure*}[h]%
\begin{minipage}{.91\textwidth}
\centering
\fbox{\parbox{.85\textwidth}{\centering
\subfloat{
\includegraphics*[width=.25\textwidth]{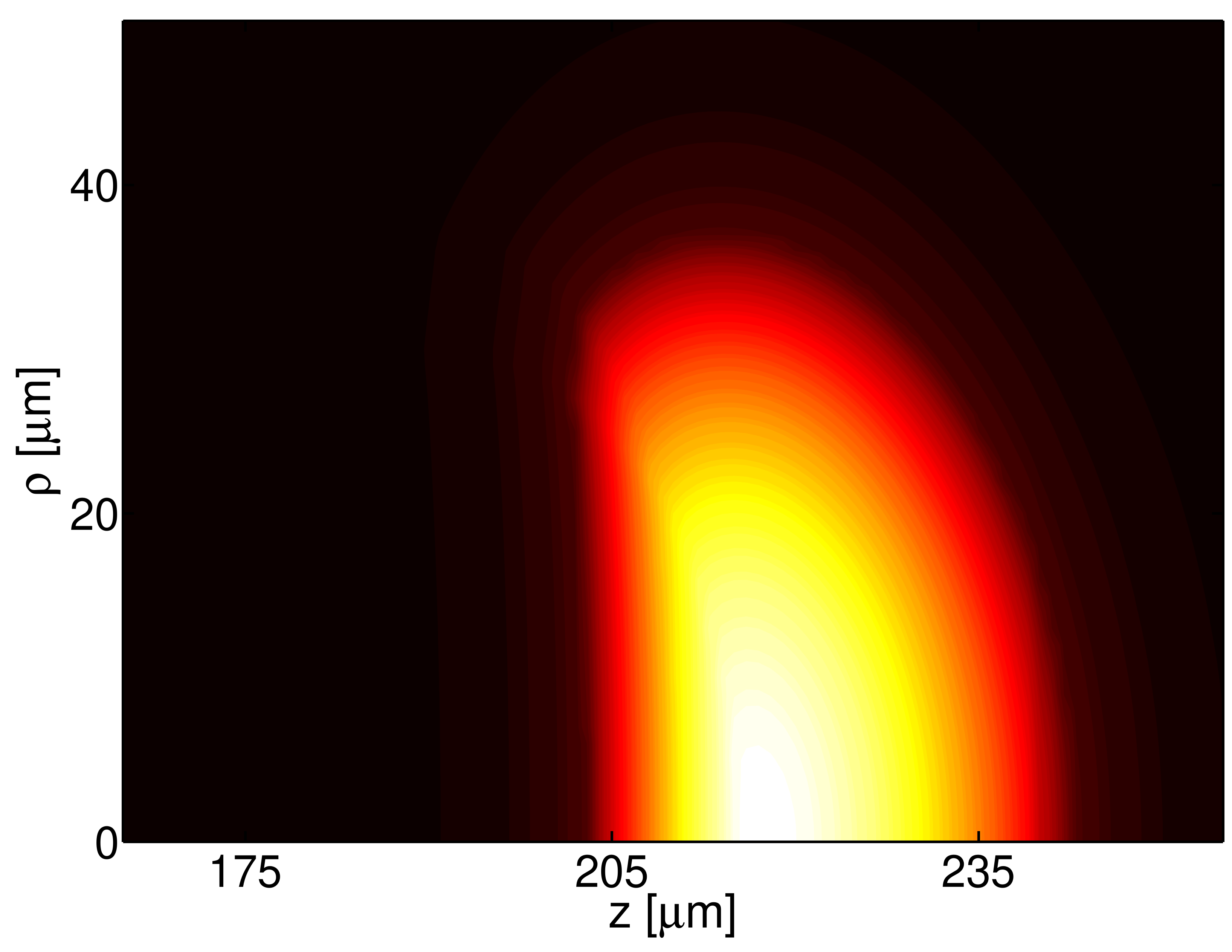}}
\subfloat{
\includegraphics*[width=.25\textwidth]{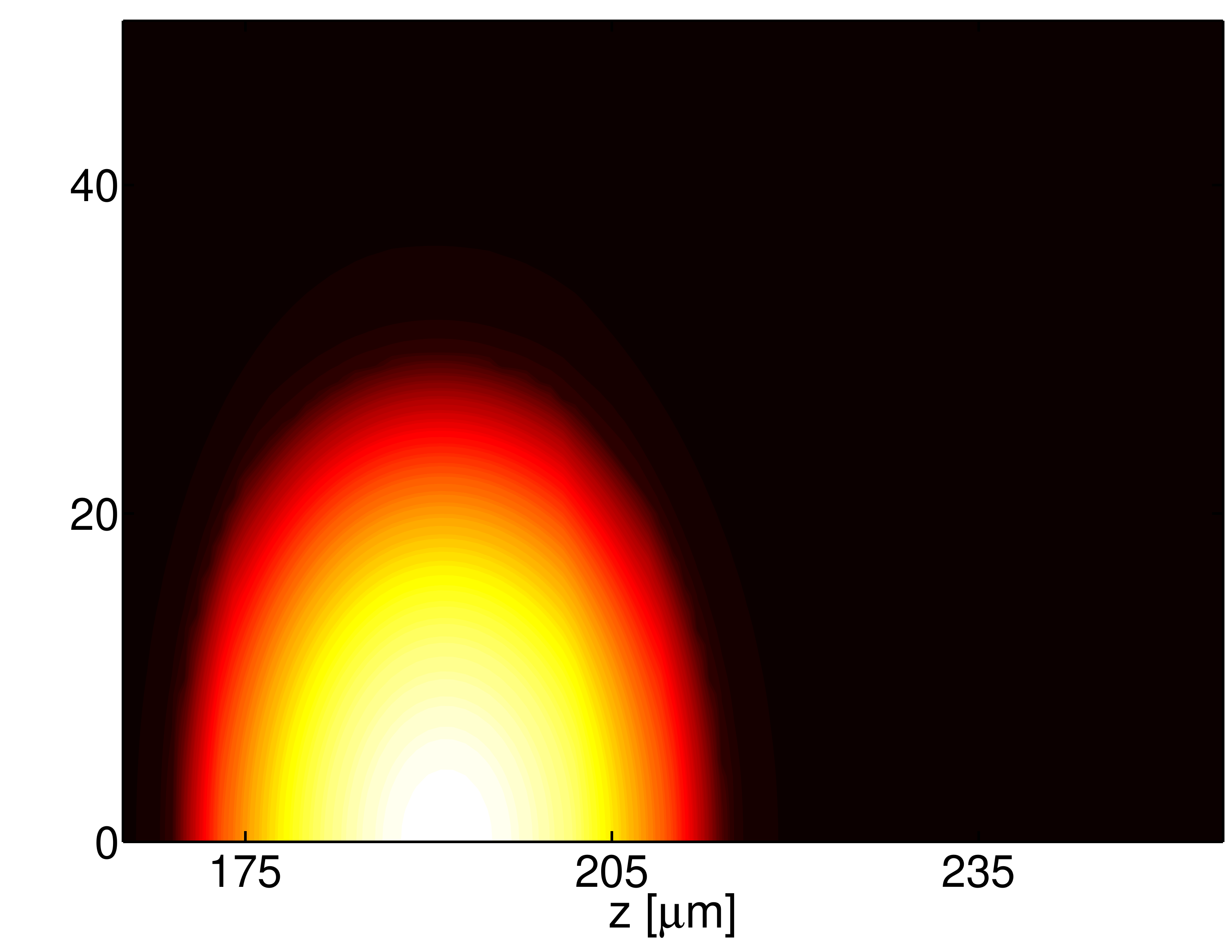}}
\subfloat{
\includegraphics*[width=.25\textwidth]{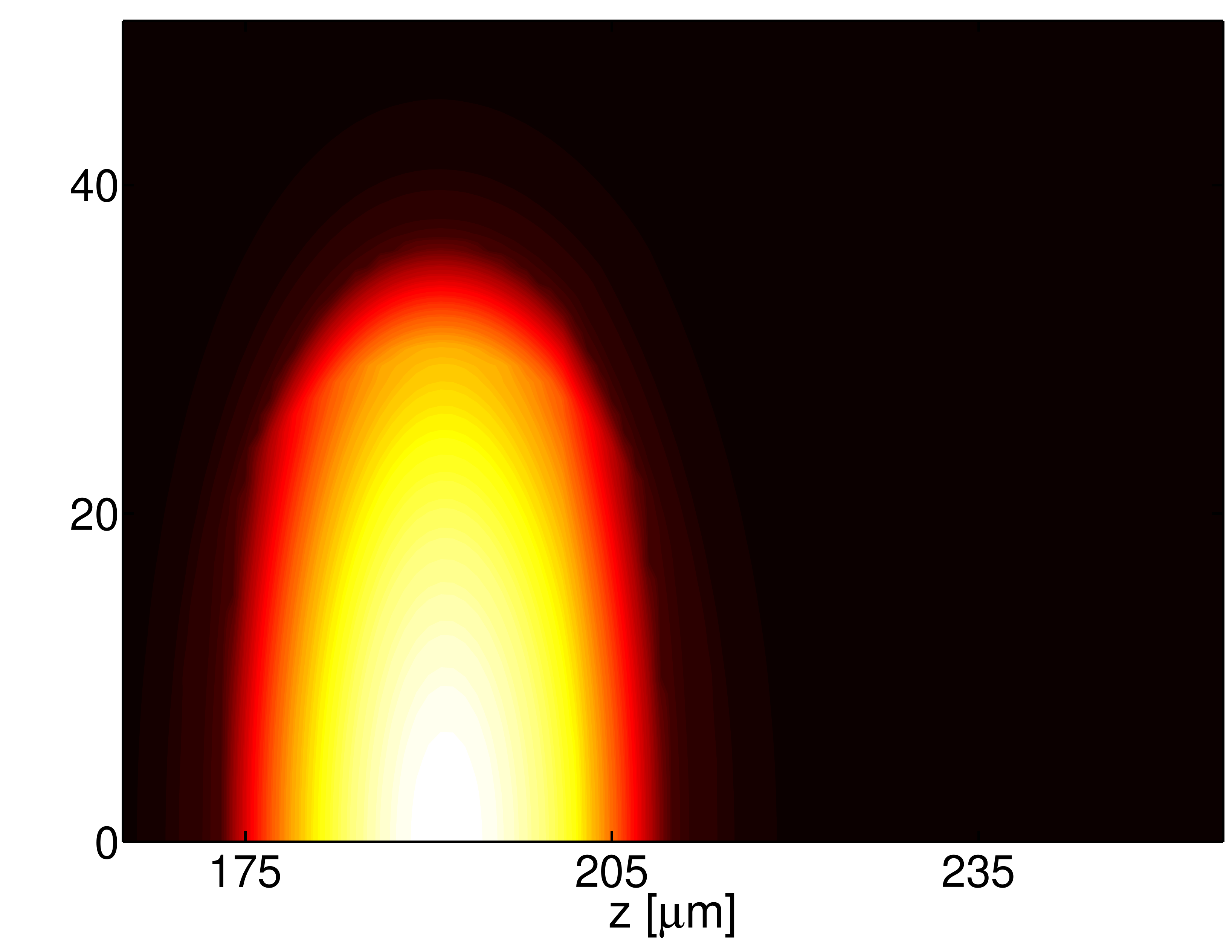}}\\[-2.4mm]
\subfloat{
\includegraphics*[width=.37\textwidth]{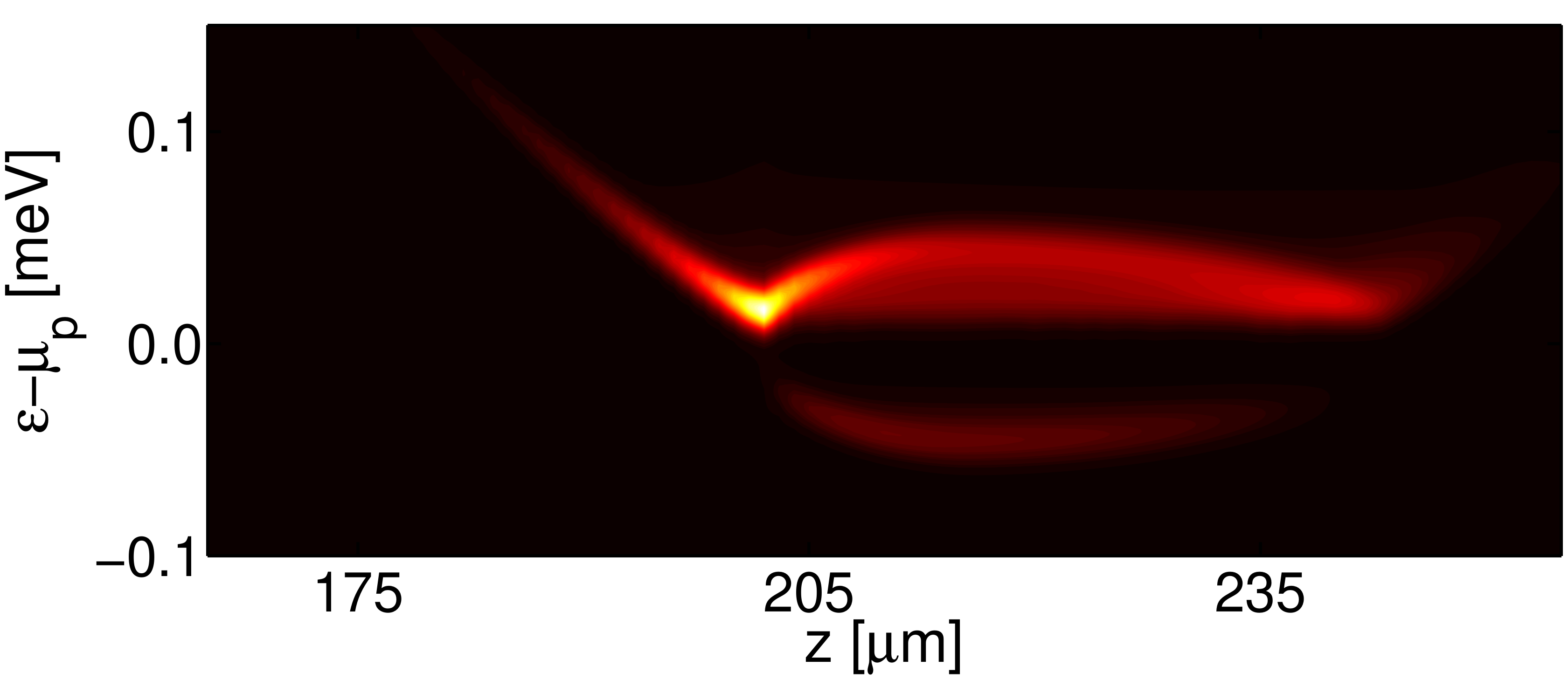}}
\subfloat{
\includegraphics*[width=.37\textwidth]{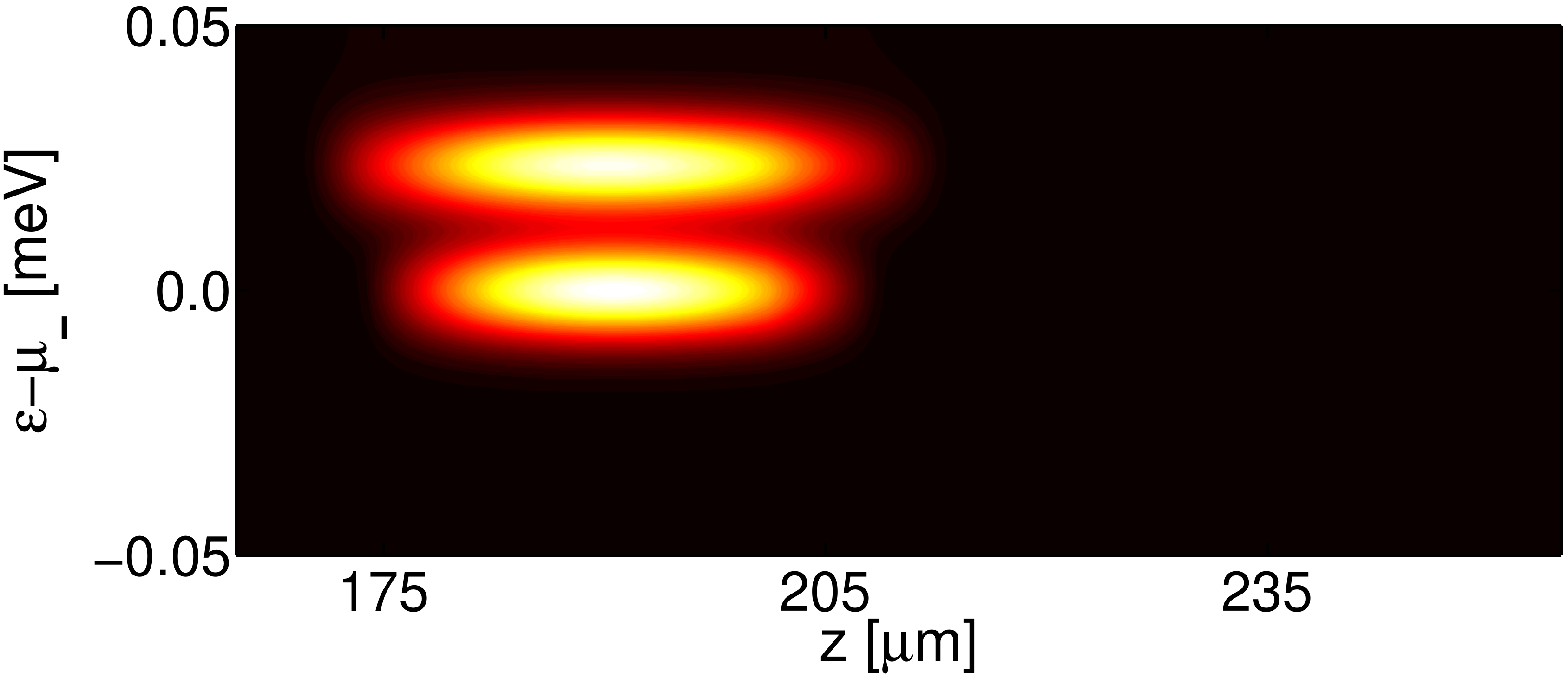}}
}}
\vspace*{-.5mm}

\fbox{\parbox{.85\textwidth}{\centering
\subfloat{
\includegraphics*[width=.25\textwidth]{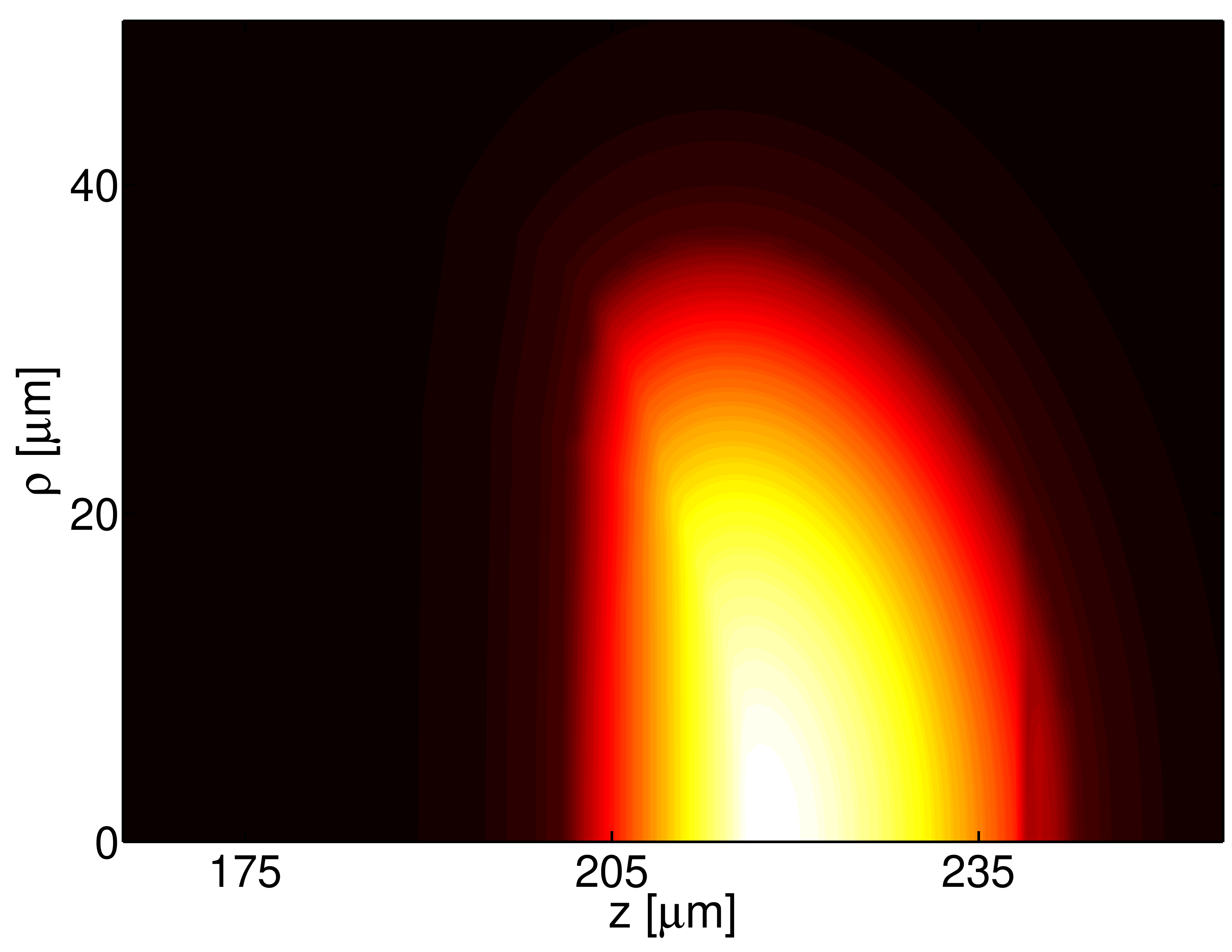}}
\subfloat{
\includegraphics*[width=.25\textwidth]{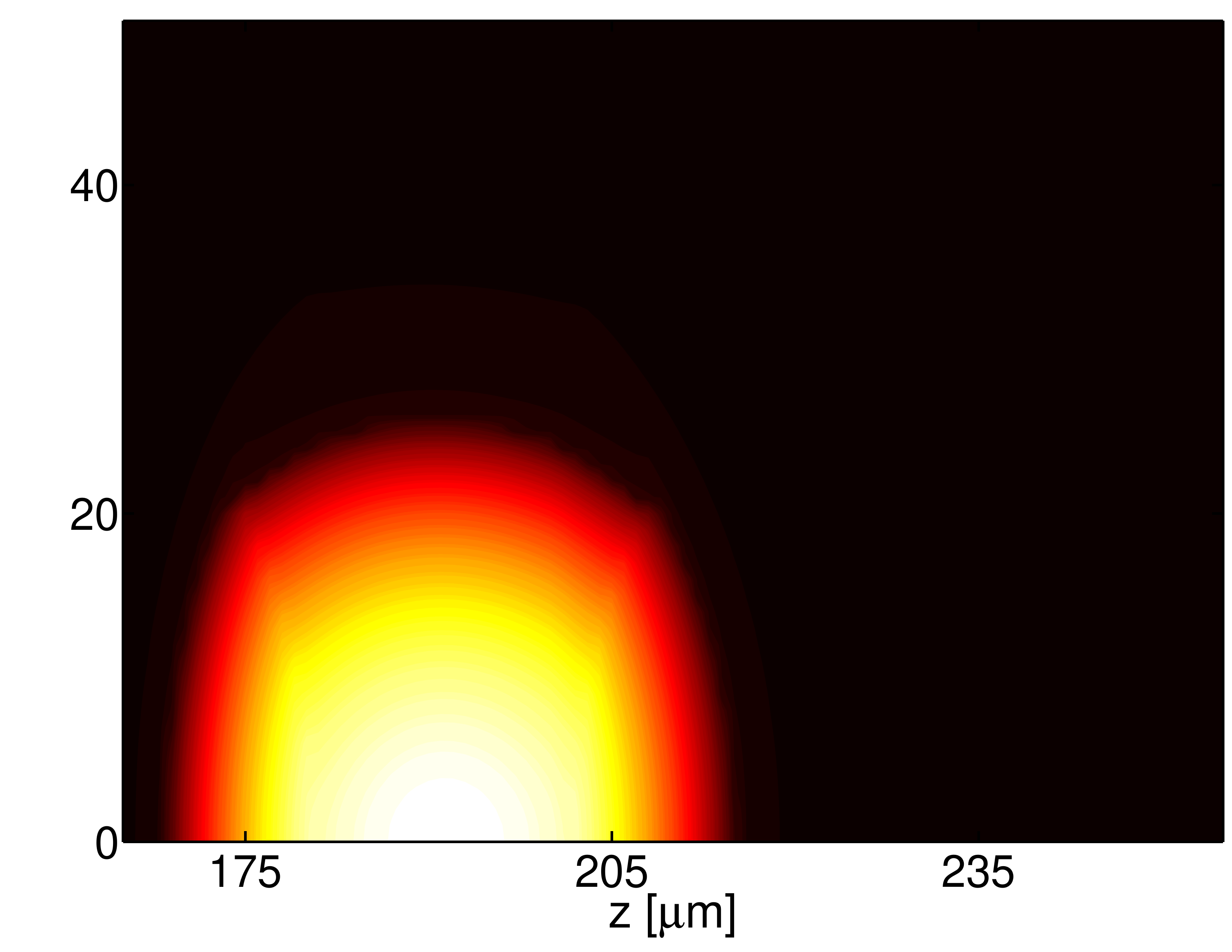}}
\subfloat{
\includegraphics*[width=.25\textwidth]{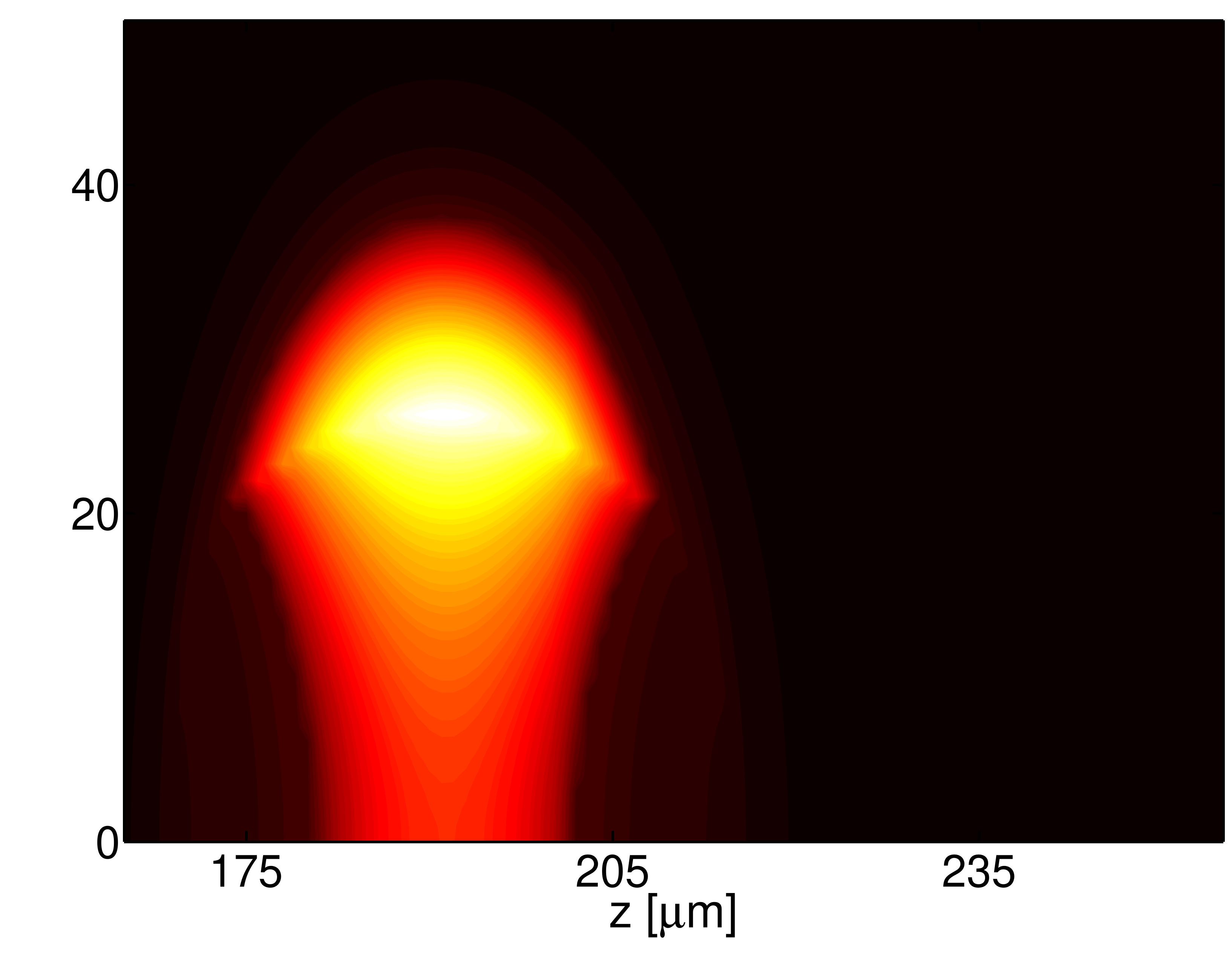}}\\[-2.4mm]
\subfloat{
\includegraphics*[width=.37\textwidth]{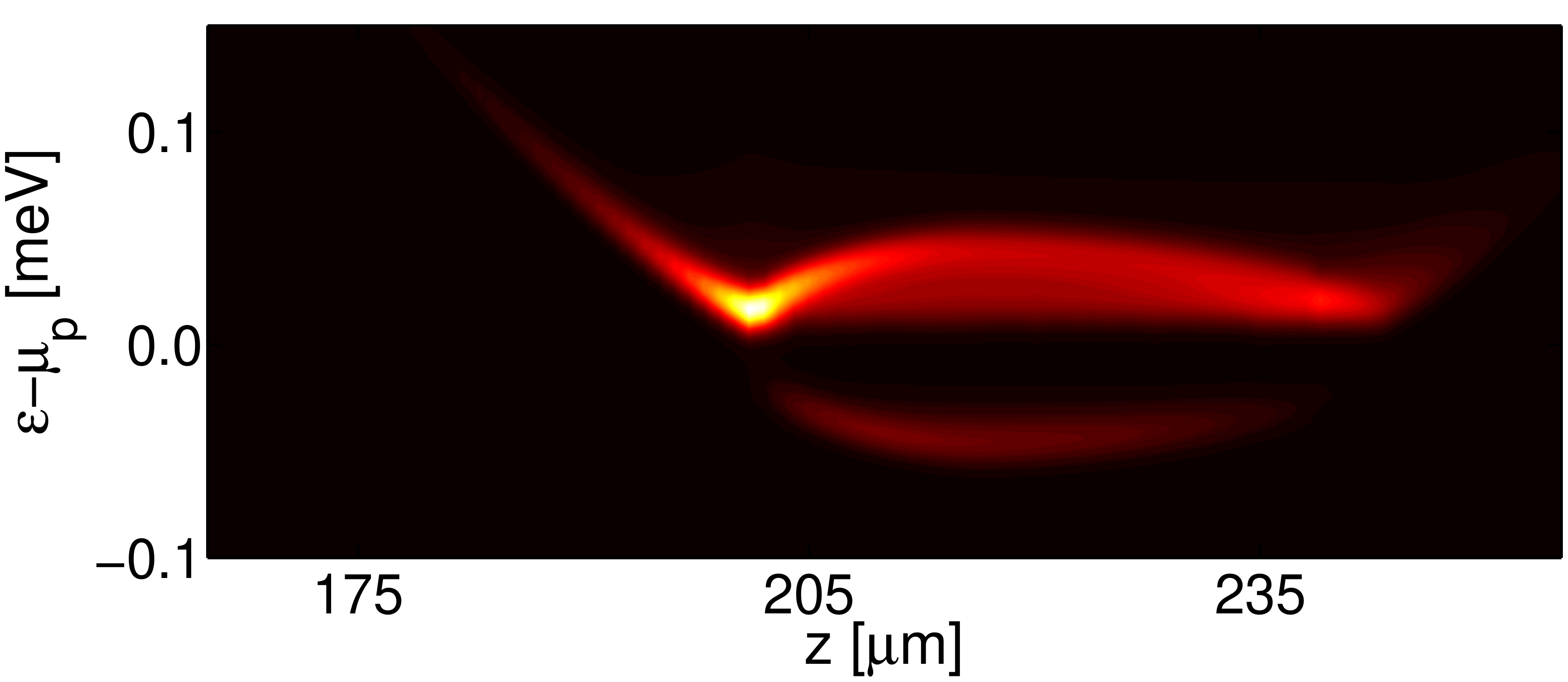}}
\subfloat{
\includegraphics*[width=.37\textwidth]{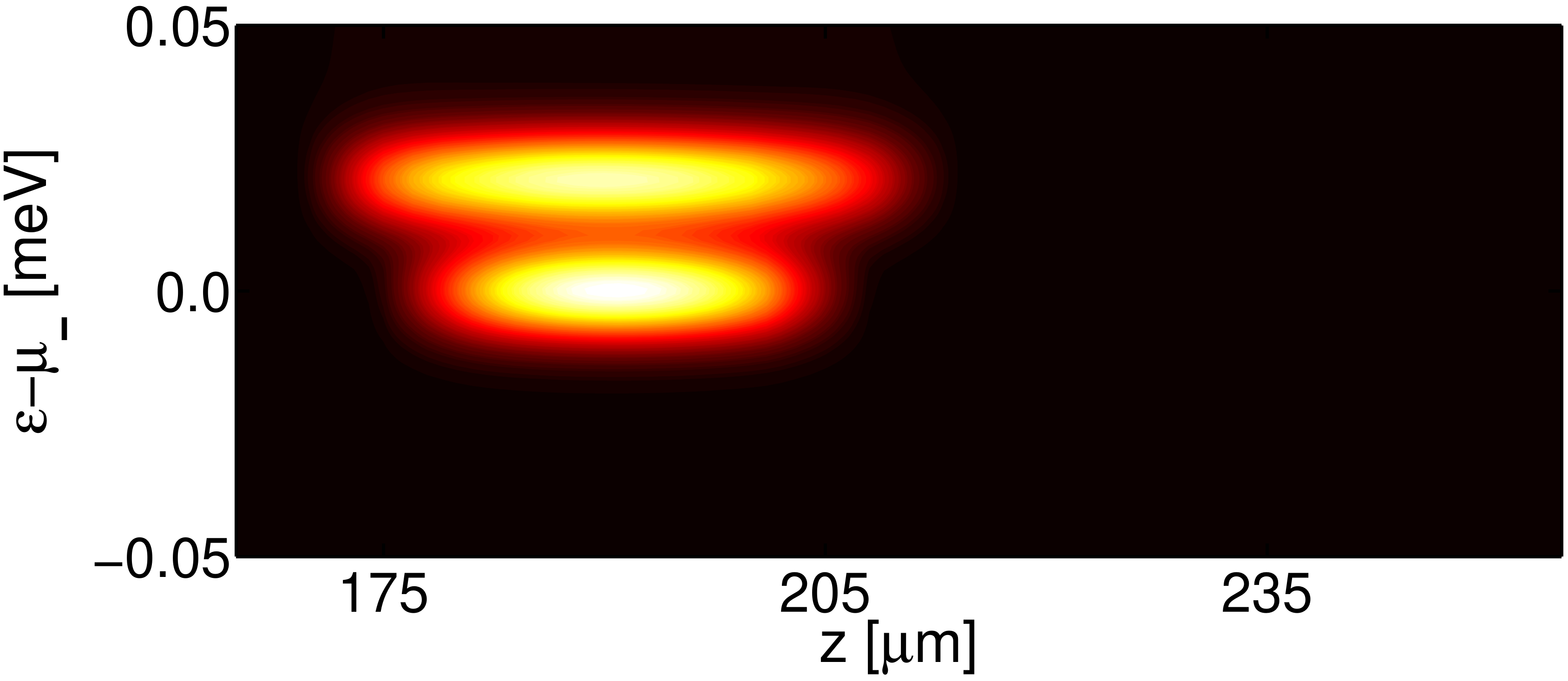}}
}}
\vspace*{-.5mm}

\fbox{\parbox{.85\textwidth}{\centering
\subfloat{
\includegraphics*[width=.25\textwidth]{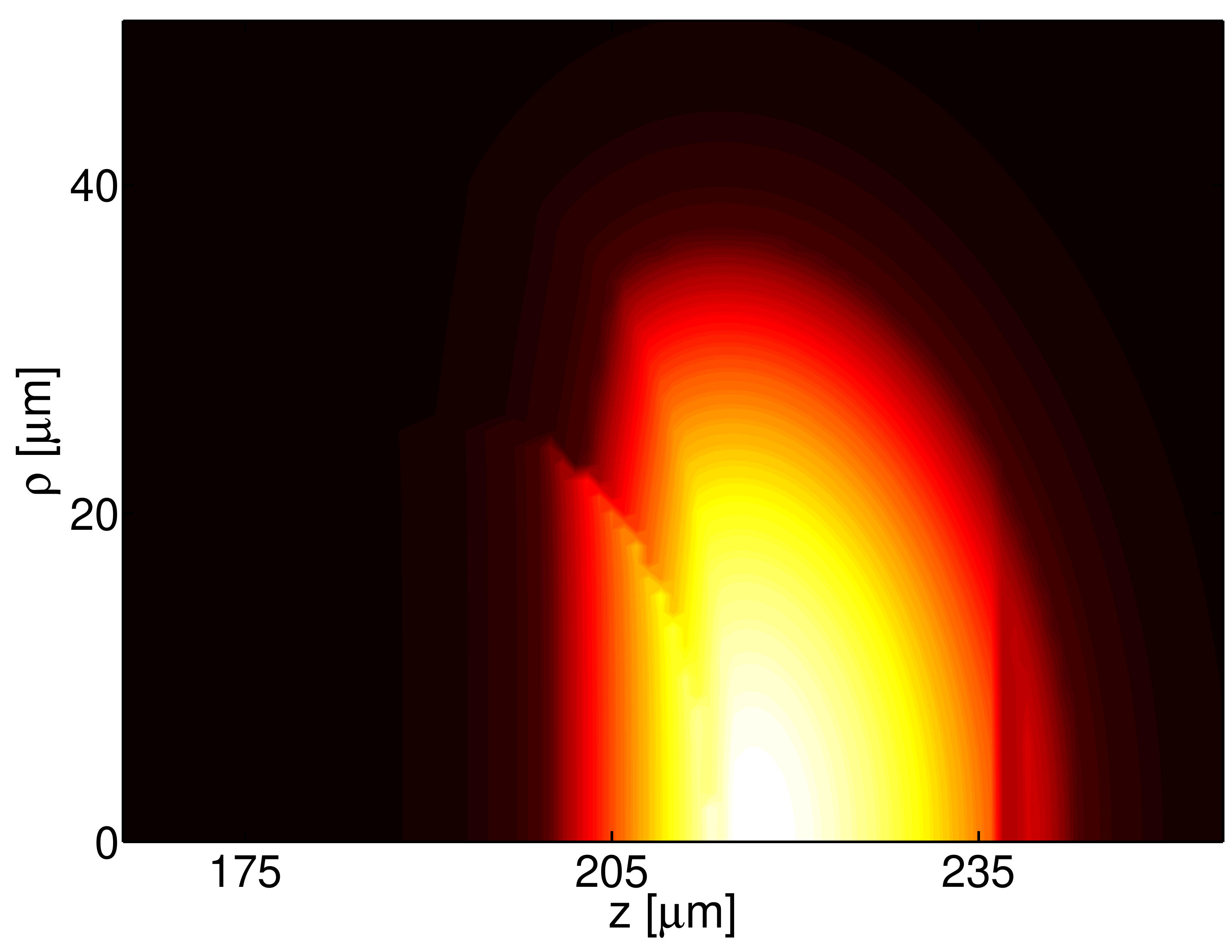}}
\subfloat{
\includegraphics*[width=.25\textwidth]{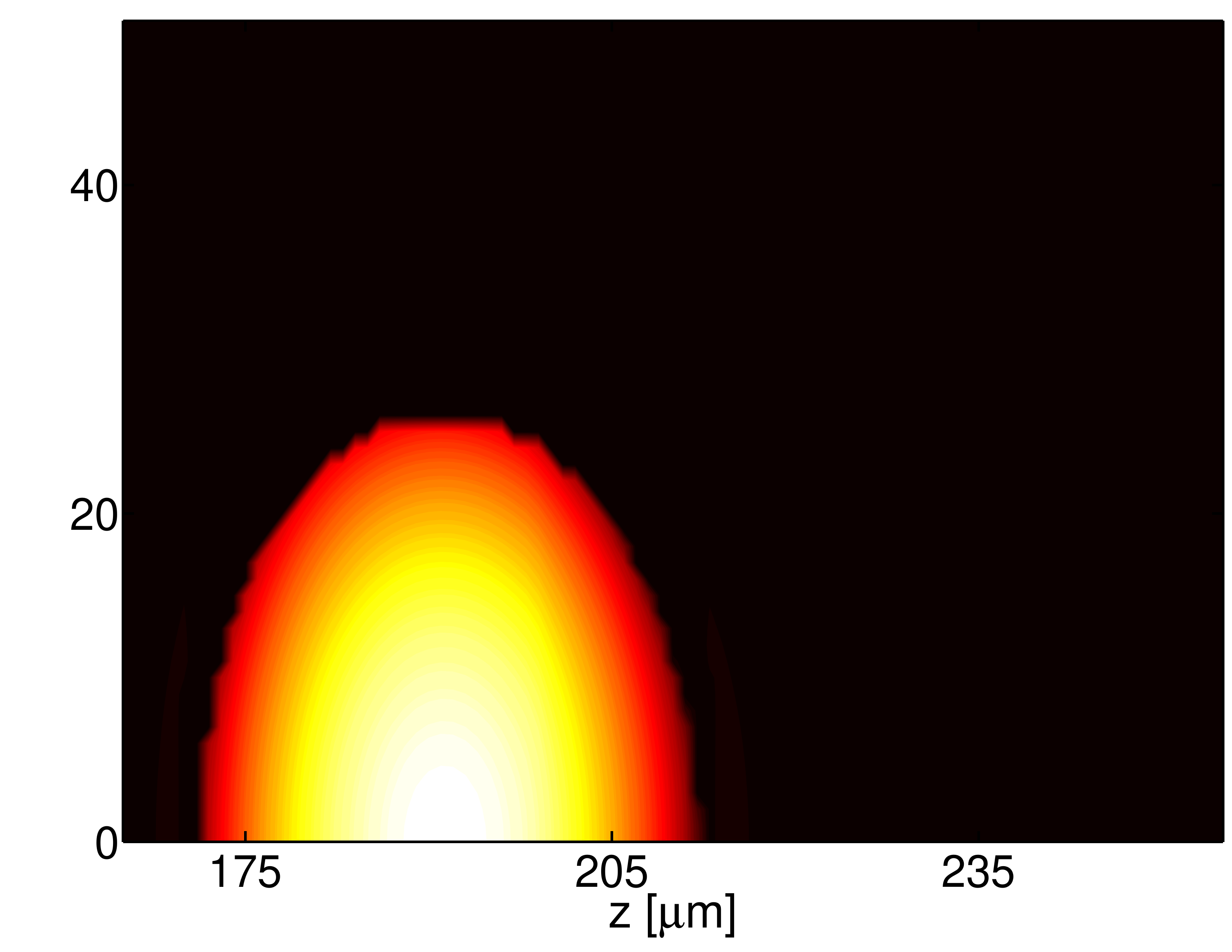}}
\subfloat{
\includegraphics*[width=.25\textwidth]{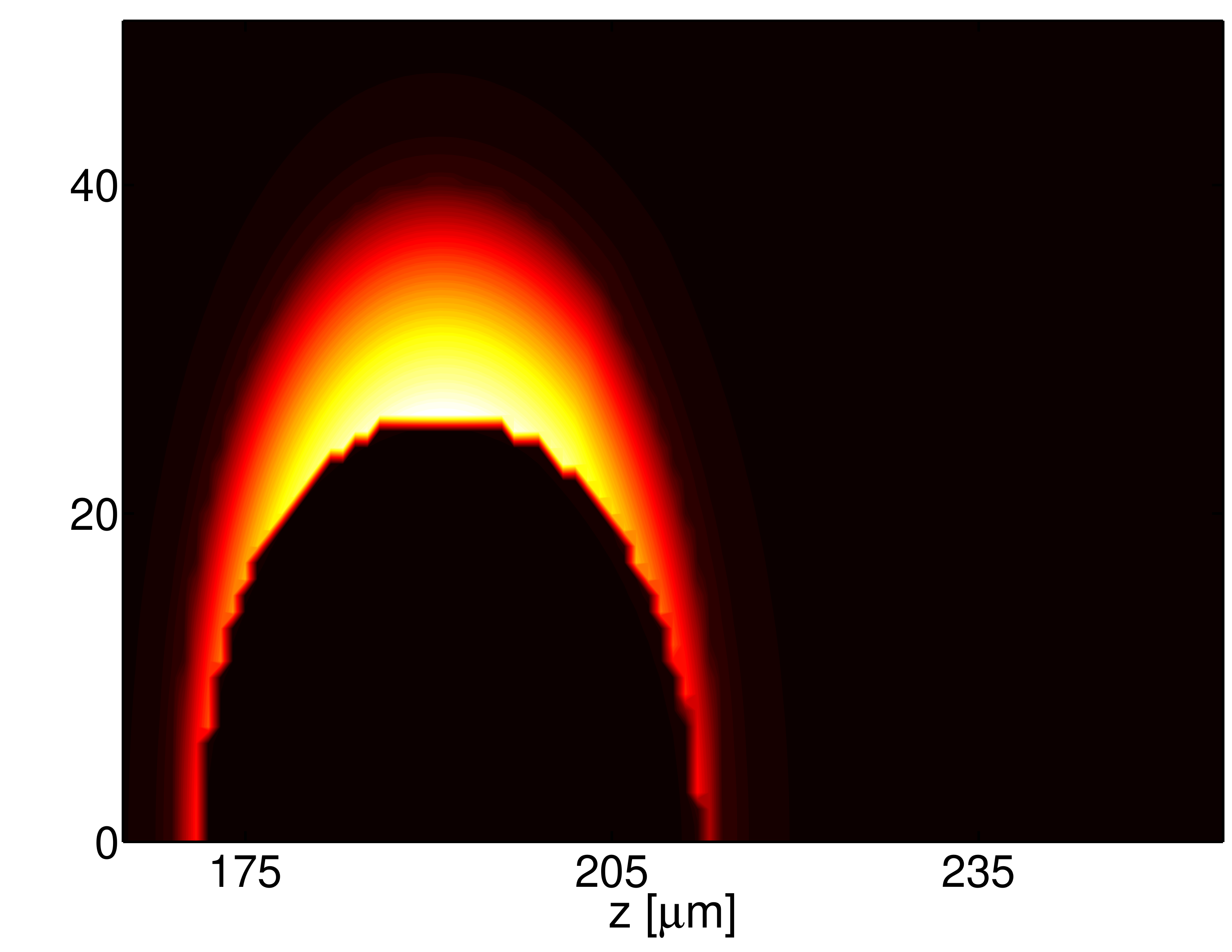}}\\[-2.4mm]
\subfloat{
\includegraphics*[width=.37\textwidth]{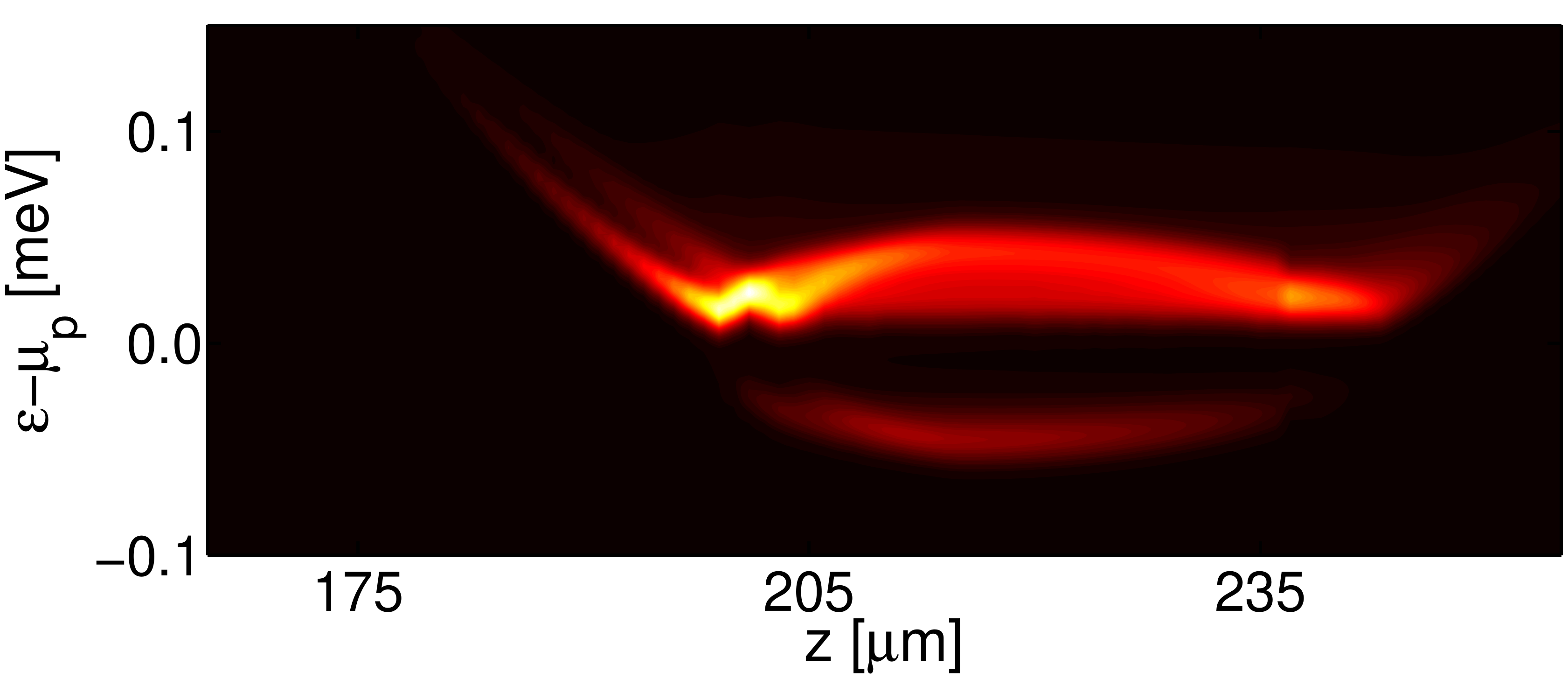}}
\subfloat{
\includegraphics*[width=.37\textwidth]{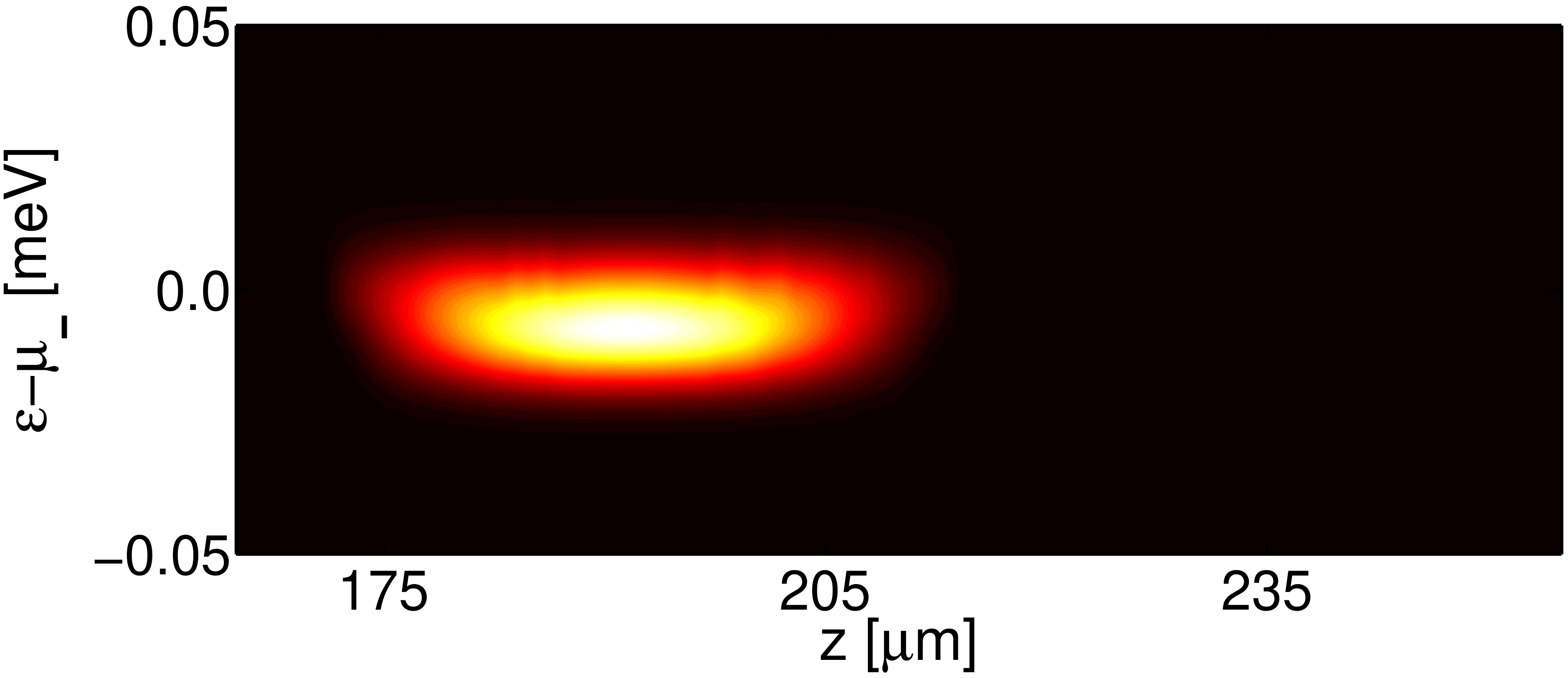}}
}}
\vspace*{-.5mm}

\caption{Three different subsets of exciton densities and luminescence spectra for para-, ortho(+)- and ortho($-$)excitons for $T=\unit{0.7}{\kelvin}$ and a total particle number of $N_p=10^{10}$ paraexcitons and $N_-=N_{+}=10^{10}$ orthoexcitons in the trap. Each subset contains 5 images with the exciton densities in the first row (from left to right para-, ortho(+)- and ortho($-$)excitons) and the spectra for the para- (left) and orthoexcitons (right) in the second row. Taking $h_{+-}$ given by Shumway and Ceperley \cite{SC01} as $h_{SC}$ the subsets represent from top to bottom: (i) $h_{+-}=h_{SC}/5$; (ii) $h_{+-}=h_{SC}/3$; and (iii) $h_{+-}=h_{SC}$ while taking all other interaction strengths as given in \cite{SC01}.}
\label{fig:specanddens}
\end{minipage}
\end{figure*}


The results of the numerical calculations are shown in Fig.\ \ref{fig:specanddens}. The densities are visualized as contour plots including thermal and condensed excitons using cylindrical coordinates ($\rho$,z). The densities are independent of $\varphi$ due to the trap geometry. The thermal excitons form the darker rim of the bright condensate spots. The second rows show the luminescence spectrum of the paraexcitons and the combined spectrum of the orthoexcitons. The latter contains the condensate signals, i.e., the first term on the r.h.s of (\ref{eqn:spectrum}), assuming $S(\mathbf{k})$ is a $\mathbf{k}$-independent constant. 

At a temperature of $T=\unit{0.7}{\kelvin}$, all three exciton species show Bose--Einstein condensation. The condition for phase separation of the orthoexcitons is approximately $h_{+-}^2>h_{++}h_{--}$ \cite{HS96} and becomes in the considered system  $h_{+-}>h_{++}$ since $h_{++}=h_{--}$. In case (i) where $h_{+-}$ is considerably smaller than $h_{++}$ the ortho-condensates are nearly completely mixed. Due to the different minimum positions of the respective external potentials, the thermal paraexcitons are pushed aside by the combined orthoexciton densities. 
The spectrum of the orthoexcitons is dominated by two condensate peaks at the chemical potentials $\mu_-=\unit{-8013}{\micro\electronvolt}$ and $\mu_+=\unit{-7990}{\micro\electronvolt}$. In case of the paraexcitons, only the zero-phonon spectral line contributes. It reveals the existence of a condensate via the flat bottom at $\varepsilon-\mu_p=0$ \cite{SS10,SSSKF10,siegfrieddiplom}.

Increasing $h_{+-}$ to $h_{SC}/3$ yields the results shown in case (ii). Here $h_{+-}$ is slightly smaller than $h_{++}$ and the condensates are still mixed in a wide area. However, a starting of the separation can be observed. Nevertheless, the spectra for the para- and orthoexcitons as well as the paraexciton density are not changed qualitatively with respect to case (i).

In case (iii) the condition for phase separation, $h_{+-}>h_{++}$, is fulfilled and the ortho-condensates form a ball-and-shell structure with finite overlap. Incidentally, the difference in their chemical potentials is smaller than the spectral resolution so that the combined ortho-spectrum gives no evidence of the phase separation. However, in the region of overlapping ortho-condensates we find a noticeable depletion of paraexcitons, which results in a W-shaped distortion of the para-spectrum when compared to case (i) or (ii). 

When the interaction strength $h_{+-}$ is further increased, no qualitative changes with respect to case (iii) are found. For $T>0$ there are no pure ($+$) or ($-$) phases \cite{VS} and the spectral features described in case (iii) remain.


\section{Conclusion and outlook}

Our simulations for the experimentally relevant example cuprous oxide have shown that at finite temperatures a possible phase separation of excitonic condensates may not be reflected in their combined luminescence spectrum. However, in the case of cuprous oxide, the single spectral line of the direct paraexciton decay may consitute a rather sensitive probe for the spatial structure of the orthoexciton density distribution. Specific distortions in the paraexciton spectrum would provide an experimental footprint of a phase separation of orthoexciton condensates. If detected, the interaction strength of the (+) and ($-$) species would satisfy the relation $h_{+-}>h_{++}$.

Omitting the Thomas--Fermi approximation or one of the other approximations used, might lead to results that quantitatively differ from the ones presented here. However, the above conclusions are mainly based on the existence of a finite overlap of separated condensates and should remain valid even if the spectra and densities are calculated beyond the used approximations. Also the calculations for the luminescence spectrum could be enhanced using more advanced approaches \cite{RFH08}. Further research is required to address these issues.

\begin{acknowledgement}
We would like to thank G.\ Manzke and W.-D.\ Kraeft (Rostock), and A.\ Alvermann (Cambridge) for many fruitful discussions.
This work was supported by the Deutsche Forschungsgemeinschaft via Collaborative
Research Center SFB 652, projects B1 and B5.
\end{acknowledgement}

\end{document}